\documentclass[referee,fleqn,usenatbib]{mnras}
\usepackage{mathptmx}
\usepackage{graphicx}	
\usepackage{amsmath}	
\usepackage{amssymb}	
\title[Second order lensing expansion]{Second order singular pertubative theory for gravitational lenses.}
\author[Alard, C.]{Alard, C.
\\
IAP, 98bis boulevard Arago Paris \\}
\date{July 5th 2016}
\pubyear{2015}
\begin{document}
\label{firstpage}
\pagerange{\pageref{firstpage}--\pageref{lastpage}}
\maketitle
%
\begin{abstract}
The extension of the singular perturbative approach to the second order is presented in this paper. The general
expansion to the second order is derived. The second order expansion is considered as a small correction to the
first order expansion. Using this approach it is demonstrated that the second order expansion is reducible to a first
order expansion via a re-definition of the first order pertubative fields. Even if in practice the second order correction
is small the reducibility of the second order expansion to the first order expansion indicates a degeneracy problem.
In general this degeneracy is hard to break. A useful and simple second order approximation is the thin source
approximation which offers a direct estimation of the correction. The practical application of the corrections derived in this
paper are illustrated by using an elliptical NFW lens model. The second order pertubative expansion provides a noticeable
improvement, even for the simplest case of thin source approximation. To conclude it is clear that for accurate modelisation
of gravitational lenses using the perturbative method the second order perturbative expansion should be considered. In particular
an evaluation of the degeneracy due to the second order term should be performed, for which the thin source approximation
is particularly useful.
\end{abstract}
\begin{keywords}
gravitational lensing: strong
\end{keywords}
\section{Introduction}
The singular perturbative method is a non parametric approach to gravitational lenses offering a direct relation between
the description of the lens and the observations. The direct relation between the lens and the data minimize the
degeneracy problems generally encountered in gravitational lenses modeling (see for instance  ~\cite{Saha2006},
~\cite{Wucknitz2002}, ~\cite{Chiba2002}). 
In a series of papers ~\cite{Alard2007} ~\cite{Alard2009} ~\cite{Alard2010} the first order singular perturbative method was considered.
Let's first recall the basics of the first order method.
 We consider a perturbation of the perfect ring situation. A perfect ring is obtained
when a point source is at the center of a circular potential. The images of the central
point source is an infinity of points situated on a circle. The radius of this circle
is the Einstein radius associated with the circular potential. For simplicity the Einstein radius
is reduced to unity by adopting a proper set of distance units. The introduction of a non circular 
perturbation to the circular potential results in the breaking of the circle with the consequence
that the central point has now a finite number of images in the vicinity of the circle. In practice
the source itself is not reduced to a point but has a finite size which is of the order of the potential
perturbation. Additionally the source may not be exactly at the center of the circular potential
and as a consequence has an impact parameter which is also of the order of the potential perturbation which we
call $\epsilon$, with $\epsilon \ll 1$. Using polar coordinates ($r$,$\theta$) in the lens plane, the potential reads:
\begin{equation}
\phi(r,\theta)=\phi_0(r)+\epsilon \psi(r,\theta)
\label{pot_def}
\end{equation}
The lens equation relating the lens plane coordinates ${ \bf r}$ to the source plane coordinates ${\bf r_S}$ reads:
\begin{equation}
{\bf r_S} = {\bf r} -\nabla \phi
\label{lens_eq}
\end{equation}
The radial deviation from the circle is of the same order as the potential perturbation, thus
$r=1+\epsilon dr$. By inserting  Eq. ~\ref{pot_def} in the lens equation and developing to the first order
in $\epsilon$ we obtain a set of equations already presented in \cite{Alard2007}
\begin{equation}
 {\bf r_S} = \left(\kappa_2 \ dr-f_1 \right) {\bf u_r} - \frac{d f_0}{d \theta }\bf {u_{\theta}}
 \label{pert_1}
\end{equation}
And:
\begin{equation}
 f_1=\left[\frac{\partial \psi}{\partial r} \right]_{r=1}  \ \ f_0=\psi(1,\theta) \ \  \kappa_2=1-\frac{d^2 \phi_0}{d r^2}
\end{equation}
Considering that the source has an impact parameter, ${\bf r_0}=(x_0,y_0)$ it is useful to re-write Eq. ~\ref{pert_1}
using the variable ${\bf r_S={\tilde r_S}+r_0}$:
\begin{equation}
 {\bf \tilde r_S} = \left(\kappa_2 \ dr-\tilde f_1 \right) {\bf u_r} - \frac{d \tilde f_0}{d \theta }\bf {u_{\theta}}
 \label{pert_1i}
\end{equation}
With:
\begin{equation}
 \tilde f_i=f_i+x_0 \cos(\theta) + y_0 \sin(\theta)  \ \ , \ i=0,1
\end{equation}
\section{Second order expansion.}
 The perturbative development of the perfect circle situation is not limited to the first
 order in $\epsilon$. The expansion may be carried out to any order. It is particularly useful
to extend the expansion to the second order to obtain higher accuracy and quality in the reconstruction
of the lens. To carry out the second order expansion we introduce an additional field in the potential:
\begin{equation}
\left\{
\begin{aligned}
\phi(r,\theta) &= \phi_0(r)+\epsilon \psi(r,\theta) \\
\psi(r,\theta) &= f_0(\theta)+f_1(\theta) (r-1) + f_2(\theta) \frac{(r-1)^2}{2} \\
 f_2&=\left[\frac{\partial \psi}{\partial r} \right]_{r=1}
\end{aligned}
\right.
\end{equation}
Inserting in Eq. ~\ref{lens_eq} and developing to second order in $\epsilon$:
\begin{equation}
{\bf \tilde r_S} = \left( \kappa_2 \ dr -\kappa_3 \frac{dr^2}{2} -\tilde f_1-f_2 dr \right) {\bf u_r} -\left(\frac{d \tilde f_0}{d \theta}+\left(\frac{d f_1}{d \theta}-\frac{d f_0}{d \theta} \right) dr \right) {\bf u_{\theta}}
\label{ordre2}
\end{equation}
$$
{\rm With:} \ \ \kappa_3=\left[\frac{d^3 \psi_0}{dr^3}\right]_{r=1} 
$$
The order of Eq. ~\ref{ordre2} in $dr$ implies that for circular source contours the equation for $dr$ is of fourth order instead of second order
for the first order expansion (\cite{Alard2007}). However in the regime $dr \ll 1$ the effect of the second order terms is only to introduce a small perturbation of the first order expansion. Consequently the $dr$ expansion reads:
%
%
%
\begin{equation}
dr=dr_1+\epsilon dr_2
\label{dr2_eq}
\end{equation}
Where $dr_1$ corresponds to the first order expansion and $dr_2$ is the second order correction. 
By re-expanding Eq. ~\ref{ordre2} to second order in $\epsilon$ using Eq. ~\ref{dr2_eq} we obtain:
\begin{equation}
{\bf \tilde r_S}=\left(\kappa_2 dr -\tilde f_1 -\kappa_3 \frac{dr_1^2}{2}  - f_2 dr_1 \right) {\bf u_r}-\left(\frac{d \tilde f_0}{d \theta}+\left(\frac{d f_1}{d \theta}-\frac{d f_0}{d \theta} \right) dr_1 \right) {\bf u_{\theta}}
\label{ordre2_dr2}
\end{equation}
It is straightforward to reduce Eq. ~\ref{ordre2_dr2} to the first order expansion Eq. ~\ref{pert_1i} by making the following substitutions:
\begin{equation}
\left\{
\begin{aligned}
f_1 & =f_1+\kappa_3 \frac{dr_1^2}{2} + f_2 dr_1 \\
\frac{d f_0}{d \theta} &= \frac{d f_0}{d \theta}+\left( \frac{d f_1}{d \theta}-\frac{d f_0}{d \theta} \right) dr_1
\end{aligned}
\right.
\label{substit_1}
\end{equation} 
Note that the second order correction to the fields presented in Eq. (~\ref{substit_1}) can be iterated. Once the fields
have been corrected new positions for the images can be estimated and used as new entries to estimate another correction
for the fields. By iterating this process a  full convergence to the second order expansion is obtained.
A useful approximation is to consider thin arcs, assuming that the size of the source is of second order, the first order
expansion is reduced to $dr_1=\frac{\tilde f_1}{\kappa_2}$. As a consequence for thin arcs the second order expansion reads:
\begin{equation}
{\bf \tilde r_S}=\left(\kappa_2 dr -\tilde f_1 -\kappa_3 \frac{\tilde f_1^2}{2 \kappa_2^2} -\frac{\tilde f_1 f_2}{\kappa_2} \right) {\bf u_r}-\left(\frac{d \tilde f_0}{d \theta}+\left(\frac{d f_1}{d \theta}-\frac{d f_0}{d \theta} \right) \frac{\tilde f_1}{\kappa_2} \right) {\bf u_{\theta}}
\label{ordre2_dr2_thin}
\end{equation}

In the thin arc approximation it is possible to derive an explicit substitution to recover the first order expansion:
\begin{equation}
\left\{
\begin{aligned}
f_1 & =f_1+\kappa_3 \frac{\tilde f_1^2}{2 \kappa_2^2} + f_2 \frac{\tilde f_1}{\kappa_2} \\
\frac{d f_0}{d \theta} &= \frac{d f_0}{d \theta}+\left( \frac{d f_1}{d \theta}-\frac{d f_0}{d \theta} \right) \frac{\tilde f_1}{\kappa_2} 
\end{aligned}
\right.
\label{substit_2}
\end{equation} 
The corrective terms in Eq. ~\ref{substit_2} could be included in $f_1$ and $f_0$ or be considered as genuine order 2 corrections. As a consequence Eq. ~\ref{substit_2} describes explicitly the degeneracy in the reconstruction of the fields. Breaking this degeneracy is
difficult since it would require information at sufficient radial distance for the same angular position $\theta$ which is very hard to find in practice. The best opportunity to break this degeneracy would be to have several sources situated at different distances and thus having
different effective Einstein radius.
\section{Practical implementation by using a numerical experiment.}
\begin{figure}
 \includegraphics[width=\columnwidth]{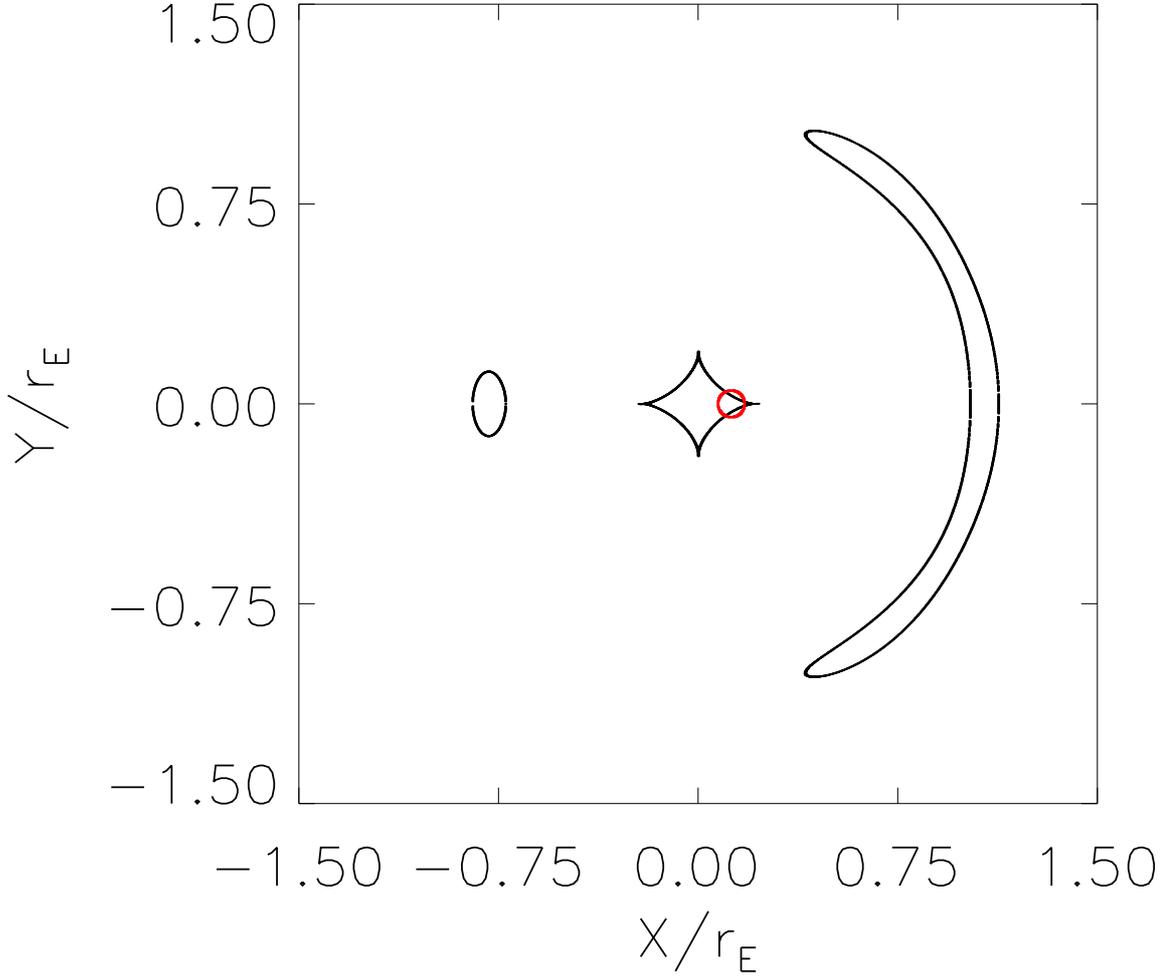}
  \caption{The source position and its associated images for the NFW lens. The images contours (black)
corresponds to the circular source contour (red). The diamond shaped curve at the center corresponds to the
caustics of the NFW elliptical lens.}
 \label{fig_1}
\end{figure}

\begin{figure}
 \includegraphics[width=\columnwidth]{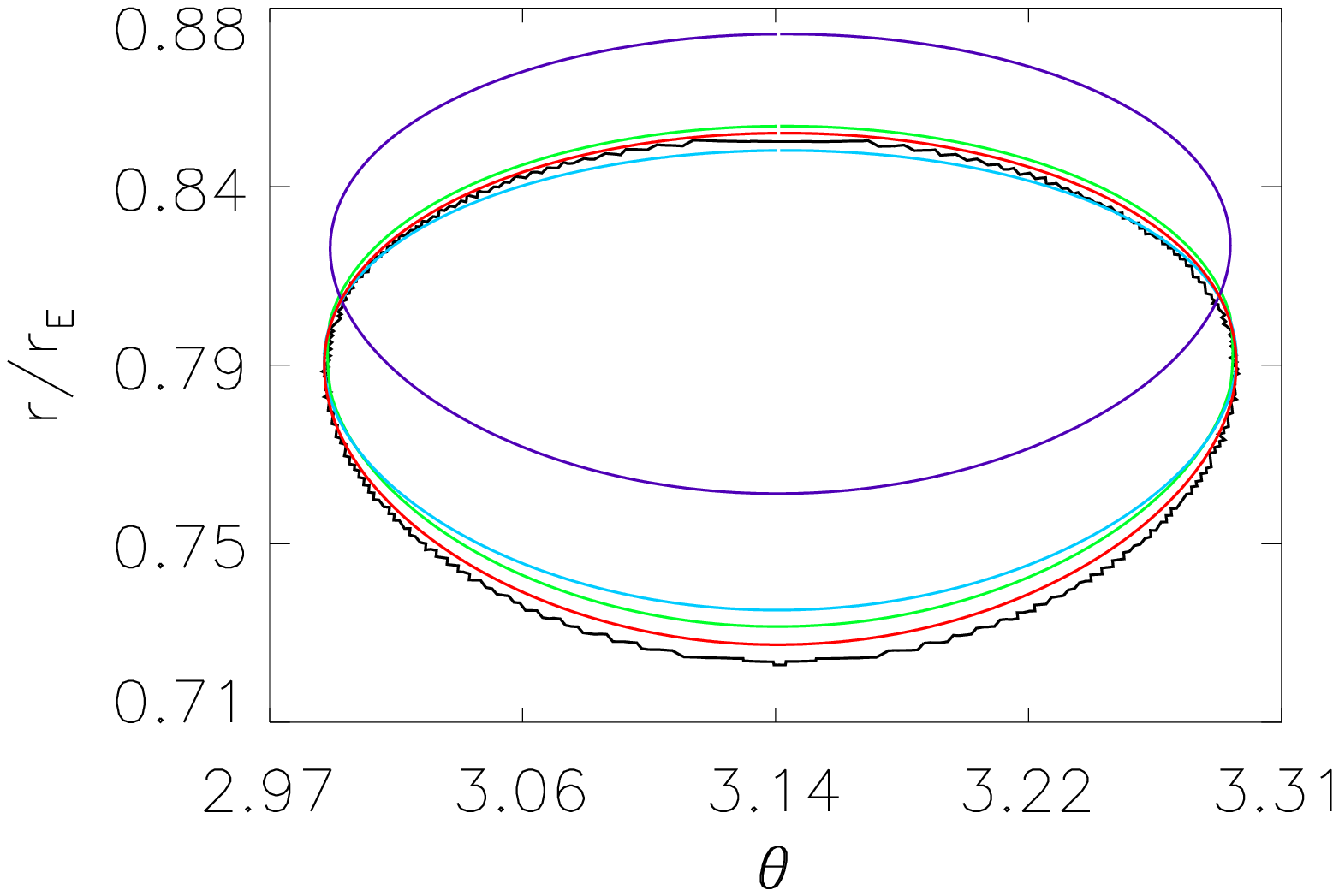}
  \caption{A detailed view of the left image (see ~\ref{fig_1}). The actual image contour (black) is super-imposed
with the first order reconstruction (blue), the second order thin source approximation (light blue), the first iteration
of the second order reconstruction (green) and the iterated second order reconstruction (red).}
 \label{fig_2}
\end{figure}

\begin{figure}
 \includegraphics[width=\columnwidth]{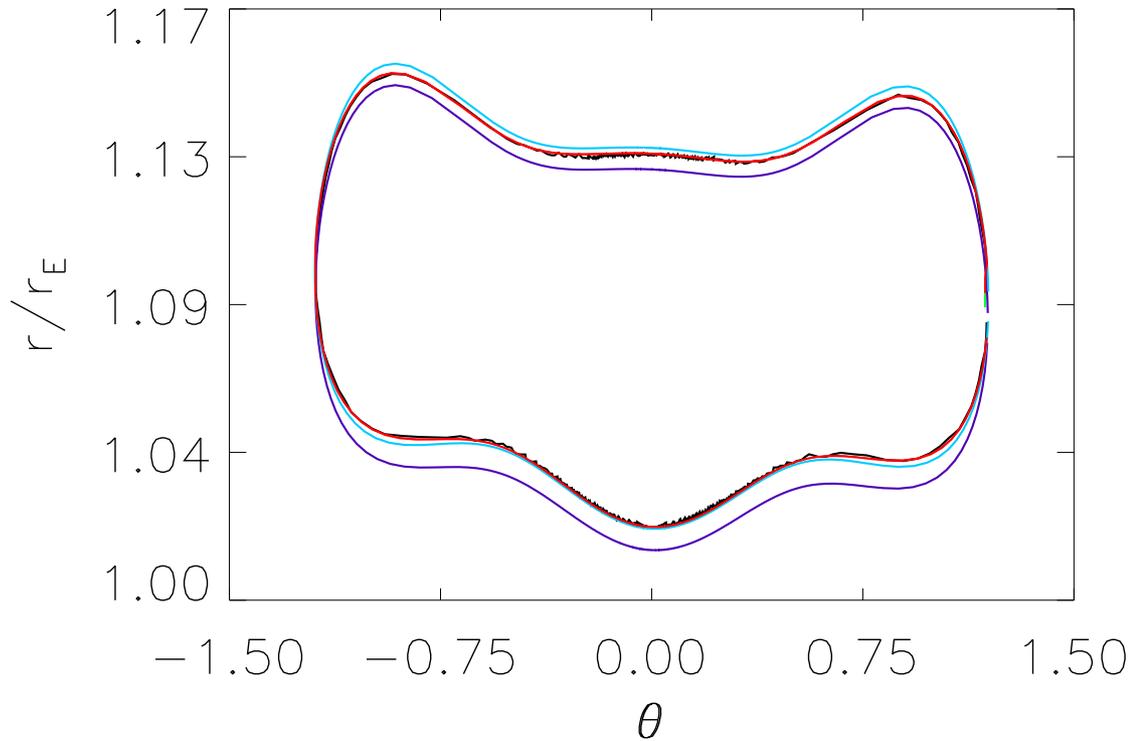}
  \caption{A detailed view of the right image (see ~\ref{fig_1}). The actual image contour (black) is super-imposed
with the first order reconstruction (blue), the second order thin source approximation (light blue), the first iteration
of the second order reconstruction (green) and the iterated second order reconstruction (red).}
 \label{fig_3}
\end{figure}

We consider the contour of a circular source situated near the caustic of a NFW halo lens. The potential for an elliptical
NFW halo is (~\cite{Meneghetti2003}):
\begin{equation}
\left\{
\begin{aligned}
 \phi(u)=\frac{1}{1-\ln(2)} g(u) \\
 u=\sqrt{\left( (1-\eta) x^2 + (1+\eta) y^2 \right)}
\end{aligned}
\right.
\label{pot_nfw}
\end{equation}
The parameter $\eta$ is related to the ellipticity of the halo. The potential normalization implies that the associated Einstein radius equal to the typical halo size, which is a common situation for gravitational lenses. The definition of the function $g(u)$ reads:
\begin{equation}
g(u)=\frac{1}{2} \ln\left(\frac{u}{2}\right)^2+
\left\{
\begin{aligned}
& \ \ \ \  2 {\rm arctan}^2 \left (\sqrt {\frac{u-1}{u+1}} \right) &  \ \ u \ge 1\\
& -2 {\rm arctanh}^2 \left (\sqrt {\frac{1-u}{u+1}} \right) & \ \ u < 1
\end{aligned}
\right.
\label{g_def}
\end{equation}
The source configuration in the potential defined in Eq. ~\ref{pot_nfw} is presented in Fig. ~\ref{fig_1} with the images of the source circular
contour. All reconstructions of the circular source contour with radius $r_0$ are performed using the first order formula (~\cite{Alard2007}) and
the modified fields defined in Eq's ~\ref{substit_1} and ~\ref{substit_2} for the second order reconstructions. The first order circular contour equation is:
\begin{equation}
\kappa_2 dr = f_1 \pm \sqrt{r_0^2-\frac{d f_0}{d \theta}^2}
\label{circ_cont}
\end{equation}
%
The results obtained in Fig. ~\ref{fig_2} indicates that the first order reconstruction is not very accurate for the left image.
 All second order expansions
provide a clear improvement in accuracy. Even the simplest second order expansion, the thin source approximation (see Eq's ~\ref{ordre2_dr2_thin} and ~\ref{substit_2}) already represents a significant improvement over the first order expansion. The first iteration of the second order expansion
(see Eq's ~\ref{ordre2_dr2} and ~\ref{substit_1}) is more accurate than the thin source approximation. Iterating the second order allows
to reach the level of accuracy corresponding precisely to the second order perturbative expansion. The results for the right image (see Fig. ~\ref{fig_3}) are similar although the first order approximation is noticeably more accurate for this image.
\section{Conclusion.}
 It is relatively simple to estimate the second order perturbative expansion as a correction of the first order expansion.
In particular the correction in the thin source limit is simple and provide a noticeable improvement over the first order
pertubative expansion. The full iterative second order correction converge to the second order perturbative correction
but in most cases provides only a small additional improvement with respect to the thin source approximation. Additionally
it is interesting to note that larger sources can always be de-composed in a number of thinner sources for which the
thin source approximation is valid. Another important issue is the problem of the degeneracy of the second order correction.
Even if in most case the correction is small the problem of the possible degeneracy of first order expansion should be addressed.
For an evaluation of the amplitude of the degenerate term the thin source approximation should be particularly useful as it offers
a direct estimation. In some particular application when the degeneracy of the second order term can be broken the full estimation of the
second order expansion should be useful. 
\end{document}